# Mid-infrared laser phase-locking to a remote near-infrared frequency reference for high precision molecular spectroscopy

B Chanteau[1], O Lopez[1], W Zhang[2], D Nicolodi[2], B. Argence[1], F Auguste[1], M Abgrall[2], C Chardonnet[1], G Santarelli[2,3], B Darquié[1], Y Le Coq[2] and A Amy-Klein[1]

[1]*Laboratoire de Physique des Lasers, Université Paris 13, Sorbonne Paris Cité, CNRS, 99 Avenue Jean-Baptiste Clément, 93430 Villetaneuse, France*

[2]*LNE-SYRTE, Observatoire de Paris, CNRS, UPMC, 61 Avenue de l'Observatoire, 75014 Paris, France*

[3]*Laboratoire Photonique, Numérique et Nanosciences, Université de Bordeaux 1, Institut d'Optique and CNRS, 351 cours de la Libération, 33405 Talence, France*

*E-mail : amy@univ-paris13.fr*

**Abstract**

We present a new method for accurate mid-infrared frequency measurements and stabilization to a near-infrared ultra-stable frequency reference, transmitted with a long-distance fibre link and continuously monitored against state-of-the-art atomic fountain clocks. As a first application, we measure the frequency of an $OsO_4$ rovibrational molecular line around 10 μm with a state-of-the-art uncertainty of $8\times10^{-13}$. We also demonstrate the frequency stabilization of a mid-infrared laser with fractional stability better than $4\times10^{-14}$ at 1 s averaging time and a linewidth below 17 Hz. This new stabilization scheme gives us the ability to transfer frequency stability in the range of $10^{-15}$ or even better, currently accessible in the near-infrared or in the visible, to mid-infrared lasers in a wide frequency range.

## 1. Introduction

With their rich internal structure, molecules can play a decisive role in precision tests of fundamental physics. They are for example now being used to test fundamental symmetries [1-3], and to measure either absolute values of fundamental constants [4] or their temporal variation [5-6]. Most of those experiments can be cast as the measurement of molecular frequencies. Ultra-stable and accurate sources in the mid-infrared (MIR) spectral region, the so-called molecular fingerprint region that hosts many intense rovibrational signatures, are thus highly desirable. MIR laser frequency stabilization has been performed for a long time using molecular references such as $CH_4$ or $OsO_4$ (see for instance [7-10]). However obtained stability is at least one order of magnitude below those of visible or near-infrared lasers stabilized to ultra-stable cavity. Moreover only a few molecular lines can be used when ultra-high accuracy is needed.

In this paper we present a new method for accurate mid-infrared laser frequency stabilization. The frequency reference is a near-infrared cavity stabilized laser continuously monitored against primary standards and the coherent frequency link between near-infrared and mid-infrared frequencies is obtained by using an optical frequency comb. Moreover we demonstrate this stabilization scheme with a remote near-infrared frequency reference transferred via an optical fibre link from a national metrological institute (NMI). This technique is thus accessible to any laboratory which can be connected to such a NMI with a fibre optical link [11].

Optical frequency combs have proven to be essential for laser frequency measurement and stabilization from the infrared to ultraviolet domain (see for instance [12]). Fractional accuracy and stability (at $10^4$ s averaging time) down to a few $10^{-16}$ are potentially reachable when the frequency reference is provided by advanced primary standards. Extension to MIR spectral domain has been demonstrated by comparing the MIR laser frequency to a very high harmonic of the comb repetition rate using sum or difference frequency generation [13-20]. Efforts have also been made towards the development of MIR frequency combs [10, 21-26].

In this paper we first describe the coherent frequency stability transfer between near- and mid-infrared frequencies around 10 µm. Then we demonstrate absolute frequency measurement of a MIR frequency with a fractional resolution of at least $4\times10^{-14}$. We also report a first application to high resolution molecular spectroscopy with a fractional uncertainty of 8x10$^{-13}$ on the line centre. Finally we present the mid-infrared laser frequency stabilization against the near-infrared frequency reference leading to state-of the art stability.

## 2. Experimental set-up

The experimental setup is shown in figure. 1. The ultra-stable optical reference located at LNE-SYRTE is a 1.54 µm fibre laser locked to a high finesse cavity. Its fractional frequency instability was measured to be lower than $2\times10^{-15}$ at 1 s and $10^{-14}$ at 100 s (after a 0.3-Hz/s drift was removed) [27]. Its frequency is measured using a fibre fs laser centered around 1.55 µm. The laser repetition rate is phase-locked to the optical reference frequency after removal of the comb frequency offset $f_0$. Fast corrections are applied to an intra-cavity electro-optic modulator (bandwidth > 400 kHz) and slower corrections to a piezo-electric transducer (PZT) controlling the laser cavity length (bandwidth ~ 10 kHz) [28]. The absolute frequency of the comb repetition rate 36$^{th}$ harmonic (9 GHz) is continuously measured against the primary standards of LNE-SYRTE, which includes an H-maser, a cryogenic oscillator and Cs-fountains [29-30]. It enables real-time measurement of the ultra-stable laser frequency drift and its correction by applying to the driving frequency of an acousto-optic modulator an opposite linear drift (with a step every ms) updated every 100 s. This makes up an ultra-stable near-infrared reference, the frequency of which is currently traceable to primary standards with a $10^{-14}$ uncertainty after 100 s.

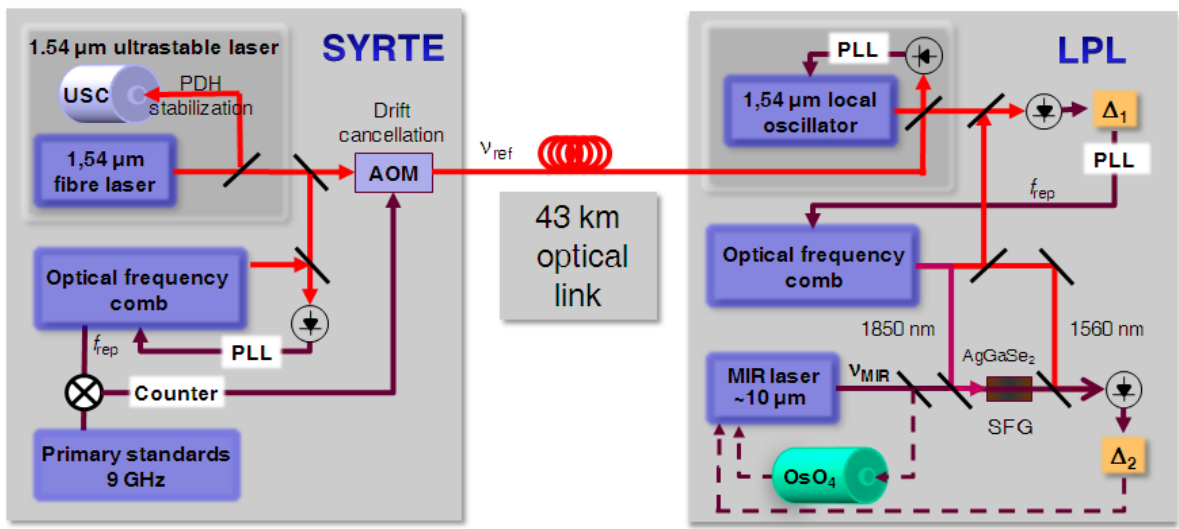

**Figure 1.** Experimental set-up. The MIR laser frequency can either be controlled with the beat-note $\Delta_2$ or the OsO$_4$ absorption signal. PLL: phase-lock loop, PDH: Pound-Drever-Hall stabilization, SFG: sum-frequency generation, AOM: acousto-optic modulator, USC: ultra-stable cavity.

This optical reference signal is transmitted to LPL through a 43-km long optical link [27]. The free-running link exhibits a propagation instability of $2\times10^{-14}$ at 1 s and around $10^{-15}$ between 100 s and 1 day. When compensated, the link instability has been measured to be roughly $10^{-15}\,\tau^{-1}$ and to reach around $10^{-18}$ after $10^3$ s (see figure 3) [27]. The frequency stability and accuracy of the reference signal are thus preserved at the LPL optical link end.

At LPL, a low-noise laser diode (free-running linewidth below 10 kHz) is phase-locked to the incoming signal with a bandwidth of 100 kHz and constitutes the local optical frequency reference $\nu_{ref}$. The repetition rate $f_{rep}$ of a 1.55 µm fibre fs laser is phase-locked to $\nu_{ref}$. To that purpose, the beat-note $\Delta_1$ between $\nu_{ref}$ and the $N^{th}$ comb mode (N ~ 780000) is used, after removal of the comb frequency offset $f_0$:

$$\nu_{ref} - N f_{rep} = \pm\Delta_1 \qquad (1)$$

Fast and slow corrections are applied to an intra-cavity electro-optic modulator and a PZT respectively as performed at LNE-SYRTE [28]. A second beat-note $\Delta_2$ compares the MIR laser frequency $\nu_{MIR}$ around 10 µm and the $n^{th}$ harmonic of the repetition rate with n≈120000:

$$\nu_{MIR} - n f_{rep} = \pm\Delta_2 \qquad (2)$$

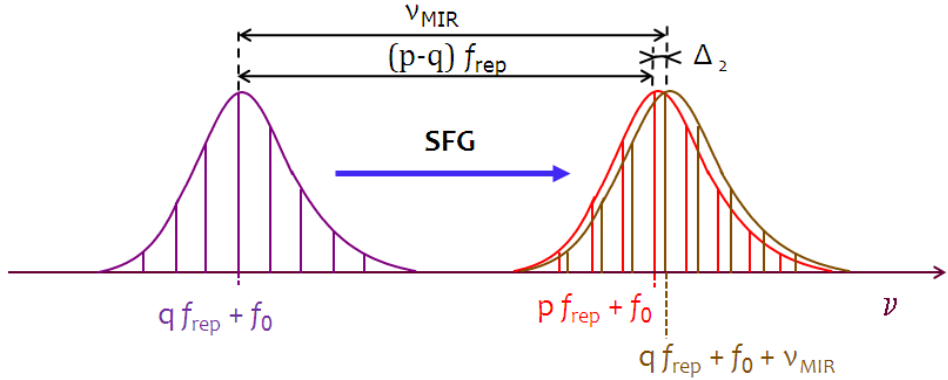

**Figure 2.** Sum-frequency of a comb output centred at 1850 nm (purple comb), of mode frequencies $q f_{rep} + f_0$ with q an integer, and the MIR laser (of frequency $\nu_{MIR}$ around 10 µm) results in a shifted comb (brown comb) centred at 1550 nm of mode frequencies $q f_{rep} + f_0 + \nu_{MIR}$. The beat-note of this shifted comb with the comb main output centred at 1550 nm (red comb), of mode frequencies $p f_{rep} + f_0$ with p an integer, can be written as $\Delta_2 = \pm\left(q f_{rep} + f_0 + \nu_{MIR} - p f_{rep} + f_0\right)$ which results in $\Delta_2 = \pm\left(\nu_{MIR} - (p-q) f_{rep}\right)$. SFG: sum-frequency generation.

This signal is generated using sum-frequency generation (SFG) of the MIR light and an additional comb output centred on 1.85 µm, generated in a non-linear fibre (figure 2) [14]. This comb output (~ 25 mW) and the MIR laser beam (~ 100 mW) are focused in a 10-mm long crystal of $AgGaSe_2$ for type I SFG. The measured efficiency is around 0.4 mW/W$^2$ and phase-matching bandwidth (for the 1.85 µm comb) is about 30 nm (~3 THz). The resulting shifted comb, centred on 1.55 µm, is combined with the 1.55 µm fs-laser output. An adjustable delay line enables to control the overlapping of the pulses in the time domain. About $10^4$ mode pairs generate the beat-note $\Delta_2$ which shows a signal-to-noise ratio of about 30 dB in a 100 kHz bandwidth. A RF tracking oscillator is phase-locked to this beat-note. Resulting from the frequency difference between two modes of the same comb, $\Delta_2$ is independent of the comb offset $f_0$.

Combining (1) and (2), the MIR laser frequency $\nu_{MIR}$ is finally obtained as:

$$\nu_{MIR} = \pm \Delta_2 + \frac{n}{N}\ \nu_{ref} \mp \Delta_1 \qquad (3)$$

with n/N roughly 0.15. The MIR frequency is thus directly linked to the near-infrared frequency reference, once the integers n and N and the signs have been determined.

**3. MIR frequency measurement and stabilisation**

To characterize the phase-coherent link between the near-infrared frequency reference and the MIR frequency, we used this setup to measure the absolute frequency of a $CO_2$ laser stabilized onto an $OsO_4$ saturated absorption line. Such a $OsO_4$-stabilized $CO_2$ laser constitutes the current state-of-the-art MIR secondary reference standard [7-8]. In this work, the $CO_2$ laser was locked either to the P(55) line of $^{190}OsO_4$ near the 10.55 µm P(16) $CO_2$ laser line or to the R(67) line of $^{192}OsO_4$ near the 10.25 µm R(20) $CO_2$ laser line. Corrections are applied to a PZT controlling the laser cavity length with a stabilization bandwidth of about 400 Hz limited by the PZT actuator. The obtained fractional frequency stability, shown in figure 3 (red circles and blue squares), is $4\times10^{-14}$ at 1 s, reaches $10^{-14}$ after 100 s of integration, and degrades at longer times due to a frequency drift of the $CO_2/OsO_4$ frequency reference. We checked that this stability was limited by the $CO_2/OsO_4$ reference since changing the $CO_2$ laser locking parameters induced a variation of the obtained stability. This stability is consistent with previous measurements obtained by comparing either two identical stabilized $CO_2$ lasers [7-8] or one of such lasers to a titanium-sapphire (Ti:Sa) frequency comb referenced to a microwave frequency [13].

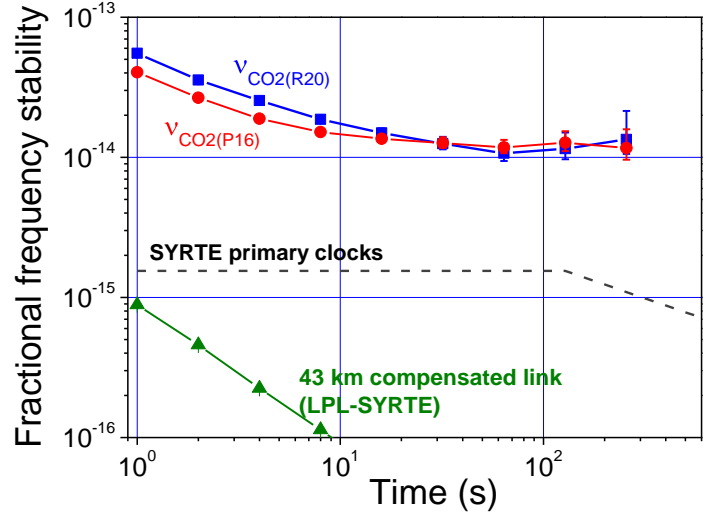

**Figure 3.** Frequency stability of a $CO_2$ laser locked on two different $OsO_4$ saturated absorption lines: the P(55) line of $^{190}OsO_4$ near the 10.55 μm P(16) $CO_2$ laser line (red circles ●) and the R(67) line of $^{192}OsO_4$ near the 10.25 μm R(20) $CO_2$ laser line (blue squares ■). The propagation instability of the compensated LPL-SYRTE fibre link (green up-triangles ▲) and the SYRTE primary standards stability (dashed line) are shown for comparison. .

This demonstrates the capability of the system to measure the stability of the best MIR frequency sources to date without any degradation. We expect the optical link and the frequency comb to contribute a few $10^{-15}$ at 1 s to the frequency instability [27-28].

The relationship of equation (3) between the frequency of the 1.54 μm LNE-SYRTE reference laser and that of the MIR laser is ensured by means of coherent phase-lock loops. Thus the accuracy of the MIR frequency measurement only depends on the uncertainty of the near-infrared frequency reference. The latter is known with an uncertainty of about $10^{-14}$ after 100 s averaging time, when only steered with the H-maser which is sufficient for this experiment. The $3\times10^{-16}$ Cs fountain accuracy [30] can ultimately be reached and then transferred from the optical reference to the MIR frequency.

As a first application to high-precision spectroscopy, we determined the absolute frequency of the P(55) line of $^{190}OsO_4$ by measuring the $OsO_4$-stabilized $CO_2$ laser frequency. Eleven measurements were performed between December 2011 and April 2012. The beat-note $\Delta_2$, the repetition rate $f_{rep}$ and the frequency $\nu_{ref}$ were counted with a gate time of 1 s. $\Delta_2$ and $\nu_{ref}$ were combined using equation (3) to calculate the $CO_2/OsO_4$ frequency. Signs in Eq. 3 and values of n and N were unambiguously deduced from $f_{rep}$ and the value $\nu_{OsO4/1999}$=28 412 648 819 596 (45) Hz of the $CO_2/OsO_4$ frequency obtained by combining two independent measurements reported in the literature [31-32]. The mean value over 600 measurements of 1 s of the $CO_2/OsO_4$ frequency gives one data point. For each data point, we correct the frequency of the LNE-SYRTE H-maser using the data published by Bureau International des Poids et Mesures (BIPM) [33]. Between each measurement the $OsO_4$ absorption cell was pumped and filled again or the whole experiment was switched off and on. We obtained $\nu_{OsO4/2012}$=28412648819588 (24) Hz where the uncertainty is the weighted 1-$\sigma$ deviation of the data points. It is -8 Hz from the value $\nu_{OsO4/1999}$ and +8 Hz from another measurement performed in 2004 with a microwave-referenced Ti:Sa frequency comb with an uncertainty of 58 Hz [13]. Within 1-$\sigma$ error bars the present result agrees with the previous measurements and confirms the very high accuracy of the measurement setup. The factor of 2 improvement of the uncertainty obtained in the measurement reported here, still limited by the molecular reference, is due to a better control of the $OsO_4$ pressure and optimization of the $CO_2$ laser locking parameters.

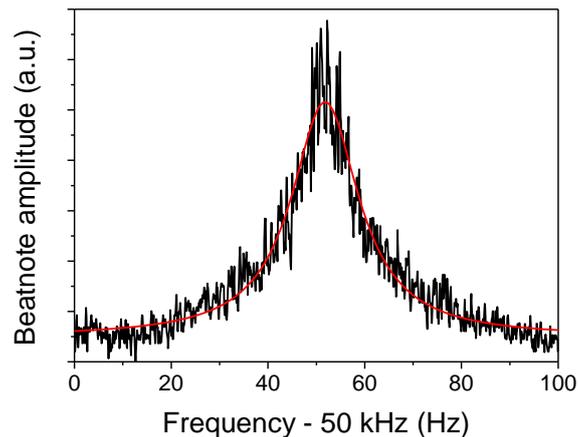

Figure 4. Down-converted beatnote of two independent $CO_2$ lasers, one stabilised on an $OsO_4$ resonance line, the other phase-locked to a comb repetition rate high-harmonic; red line is a Lorentzian fit of linewidth 17 Hz

From the previous results, we conclude that the coherent frequency chain is a viable and potentially much better alternative to an $OsO_4$ molecular transition for frequency stabilization of a MIR source. This was investigated by locking the $CO_2$ laser frequency to the optical comb: the beat-note of the free-running $CO_2$ laser with the comb repetition rate $n^{th}$ harmonic, $\Delta_2$, was phase-locked onto a stable frequency synthesizer with a 400 Hz bandwidth. The obtained $CO_2$ laser frequency stability is characterized by measuring the beat-note $\Delta_3$ with a second independent $CO_2$ laser stabilized onto $OsO_4$. Figure 4 displays this beatnote signal, fitted with a Lorentzian of linewidth 17 Hz (full width at half maximum). In the case of a Lorentzian lineshape, the contribution of each laser linewidth adds [34] and we deduce a linewidth between 8.5 and 17 Hz for each laser, the state of the art for a $CO_2$ laser [7]. Figure 5 displays $\Delta_3$'s frequency noise power spectral density (PSD) (red trace). Using the beat-note $\Delta_2$ with the local frequency comb, the frequency noise PSD of the $OsO_4$-stabilized $CO_2$ laser was also measured (figure 5, blue trace). The two PSD almost perfectly overlap as expected from efficient phase stabilization. Together with the above results, it shows that the comb-stabilized MIR laser frequency noise is at least as low as the one of the $OsO_4$-stabilized laser. The former is most probably much lower, potentially compatible with the frequency noise of the optical frequency reference which inferred PSD (including noise added by the link) is displayed in figure 5 (dotted black line). This noise level is the lowest reachable with our stabilization scheme.

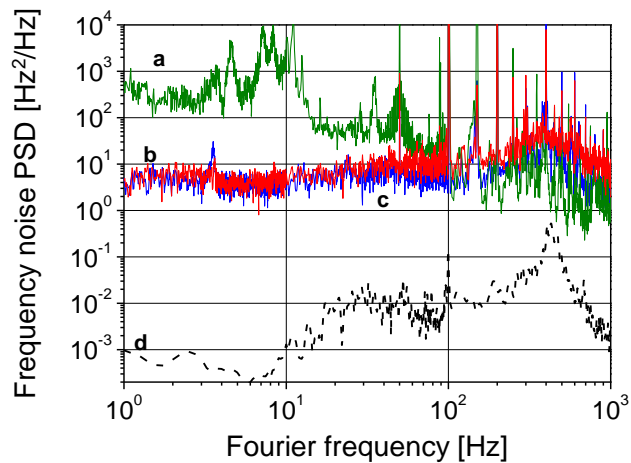

**Figure 5.** Frequency noise PSD of (a) the free-running $CO_2$ laser (green trace), (b) the beat-note between the $CO_2$ laser stabilized onto the frequency comb and an independent $OsO_4$-stabilized $CO_2$ laser (red trace), (c) the $OsO_4$-stabilized $CO_2$ laser measured with the comb (blue trace), and (d) the optical reference (dotted black line). The free-running $CO_2$ laser PSD has been measured using the beat-note $\Delta_2$ with the local frequency comb and is given for comparison.

**4. Conclusion**

We have demonstrated a coherent frequency chain linking a remote ultra-stable 1.54 µm frequency reference and a MIR source, leading to the control of the absolute MIR frequency. It uses reliable commercially available fibre-based frequency combs and an optical reference potentially available to any laboratory connected to a fibre network [11]. Stability below $4\times10^{-14}$ at 1 s was demonstrated, and we expect it to be in the $10^{-15}$ range. Using a state-of-the-art near-infrared ultra-stable laser [35] may reduce this value even further. The $3\times10^{-16}$ accuracy of the LNE-SYRTE Cs fountains is potentially within reach.

Frequency tuning of such a stabilized MIR laser source, required for high resolution spectroscopy, is achievable by scanning the near-infrared frequency referencing the comb. Tuning the frequency offset between the LNE-SYRTE optical reference and the LPL laser diode of frequency $\nu_{ref}$ (see equation (3)) would result in a tuning range of a few GHz.

This setup enables to stabilize MIR laser sources in a much wider spectral range than is currently possible using the $OsO_4$ molecular standard. With the present setup the 9-11 µm range is accessible, limited by the non-linear crystal and the central frequency of the auxiliary 1.85 µm comb output used in the sum-frequency generation. Nevertheless, it can easily be extended to the whole 5-20 µm range with proper comb spectrum and crystal optimization. Orientation-patterned GaAs would for instance ensure a wide tunability [36]. Our stabilization technique is thus particularly well-suited to quantum cascade lasers that have achievable wavelengths covering the whole MIR region [37]. Moreover, the ongoing work on dissemination of optical reference through Internet fibre networks over a continental scale [11] will eventually enable many laboratories to access an ultra-stable optical reference. Thus such ultra-stable and accurate MIR sources could benefit a very wide molecular spectroscopy community.

**Acknowledgements**

The authors acknowledge financial support from CNRS, Agence Nationale de la Recherche (ANR BLANC 2011-BS04-009-01) and Université Paris 13. Sylvain Dimanno is gratefully acknowledged for contributing to the comb driving program.